**Extending a Matrix Lie Group Model of Measurement Symmetries**

William R. Nugent, Ph.D.

College of Social Work, University of Tennessee, Knoxville, TN

orcid: 0000-0003-0811-3221

wnugent@utk.edu

**Abstract**

Symmetry principles underlie and guide scientific theory and research, from Curie's invariance formulation to modern applications across physics, chemistry, and mathematics. Building on a recent matrix Lie group measurement model, this paper extends the framework to identify additional measurement symmetries implied by Lie group theory. Lie groups provide the mathematics of continuous symmetries, while Lie algebras serve as their infinitesimal generators. Within applied measurement theory, the preservation of symmetries in transformation groups acting on score distributions ensures invariance in transformed distributions, with implications for validity, comparability, and conservation of information. A simulation study demonstrates how symmetry violations affect effect size comparability. Practical applications are considered, particularly in meta-analysis, where the standardized mean difference (SMD) is shown to remain invariant across measures only under specific symmetry conditions derived from the Lie group model. These results highlight symmetry as a unifying principle in measurement theory and its role in evidence-based research.

# Extending a Matrix Lie Group Model of Measurement Symmetries

## Introduction

Rosen (2005) re-stated Curie's symmetry principle: *"If an ensemble of causes is invariant with respect to any transformation, the ensemble of their effects is invariant with respect to the same transformation"* (p. 308). Symmetry has become a critically important principle in science. Gross (1996) observed: *"In the latter half of the 20th century symmetry has been the most dominant concept in the exploration and formulation of the fundamental laws of physics. Today it serves as a guiding principle in the search for further unification and progress"* (p. 14256). Symmetry principles are now applied not only in physics, chemistry, biology, mathematics, engineering, and computer science, but also in architecture, art and design, crystallography, nature, and even cognitive sciences such as psychology and linguistics (Kotoky, 2024). In applied measurement theory this principle can be expressed as follows: the symmetries in a transformation group acting on a frequency distribution of scores must be preserved in the transformed distribution, which may also exhibit additional invariances. This formulation parallels Rosen's (1995) statement: *"The symmetry group of the cause is a subgroup of the symmetry group of the effect"* (p. 104).

Nugent (2024) recently developed a matrix Lie group measurement model and demonstrated that several measurement symmetries are implied by this Lie-theoretic framework. Lie group theory is the mathematics of continuous symmetries. Lie algebras—examples of which are derived below—serve as the infinitesimal generators of the symmetries expressed by Lie groups. This article expands on Nugent's model by identifying additional symmetries not included in his 2024 work. Practical examples are provided, including symmetries relevant for meta-analysis. Meta-analysis is widely regarded as one of the highest levels of evidence for identifying evidence-based outcomes. Foundational to meta-analysis is the assumption that scores from different measures can be placed onto a common metric using effect sizes, such as the standardized mean difference (SMD). The author will show that the population standardized mean difference (SMD), a commonly used effect size in meta-analysis, remains invariant across different measures if and only if specific measurement conditions—contained within Nugent's (2024) matrix Lie group model—are satisfied.

The remainder of this paper is organized as follows. The next section develops the Lie group measurement framework and derivations. The following section presents a simulation study, followed by implications for meta-analysis, longitudinal, and cross-cultural research. The paper concludes with broader reflections on symmetry in measurement theory.

## Materials and Methods

### *Review and Further Development of Nugent's (2024) Model*

Following Nugent (2024), assume a set of $k$ measures of a construct of interest. Let the relationship between true scores on any measure $A$, denoted $\tau_A$, and true scores on any measure $B$, denoted $\tau_B$, from the k measures be defined as,

$$\tau_B = \gamma \cdot \tau_A + \omega \tag{1}$$

with $\gamma > 0$ and $\omega$ a real number. Similarly, assume the relationship between the standard deviation of errors of measurement is,

$$\sigma_E(Y_B) = \gamma \cdot \sigma_E(Y_A). \tag{2}$$

This transformation consists of two components:

- A symmetry transformation involving uniform scaling and translation of true scores.

- A uniform scaling transformation of the standard errors of measurement.

Both are symmetry transformations (Rosen, 1995), and the combination of (1) and (2) is also a symmetry transformation. Both transformations involve a uniform rescaling by the same parameter, $\gamma$. Another symmetry is that ratios such as,

$$\frac{\gamma \tau_A}{\gamma \sigma_E(Y_A)} = \frac{\tau_A}{\sigma_E(Y_A)} \text{ and}$$

$$\frac{\gamma \left[\mu_{\tau_A}^{P1} - \mu_{\tau_A}^{P2}\right]}{\gamma \sigma_E(Y_A)} = \frac{\left[\mu_{\tau_A}^{P1} - \mu_{\tau_A}^{P2}\right]}{\sigma_E(Y_A)},$$

where $\mu_{\tau_A}^{P1}$ and $\mu_{\tau_A}^{P2}$ are population mean true scores, are invariant, which accounts for many of the symmetries identified by Nugent (2024).

Now define the measurement vector:

$$v_A = \begin{bmatrix} \tau_A \\ \sigma_E(Y_A) \\ 1 \end{bmatrix} \tag{3}$$

where $\tau_A$ represents the true scores on measure $A$, and $\sigma_E(Y_A)$ represents the standard deviation of errors of measurement on measure $A$. The homogeneous coordinate, the 1 in the third row of the vector, anchors the vector in a two-dimensional affine space. A transformation group acts on this vector to model relationships between scores on any two measures $A$ and $B$ from the $k$ measures. The transformation matrix $\Gamma$ is:

$$\Gamma = \begin{bmatrix} \gamma & 0 & \omega \\ 0 & \gamma & 0 \\ 0 & 0 & 1 \end{bmatrix}.$$

This matrix transforms the measurement vector as follows:

$$v_B = \Gamma v_A = \begin{bmatrix} \gamma & 0 & \omega \\ 0 & \gamma & 0 \\ 0 & 0 & 1 \end{bmatrix} \begin{bmatrix} \tau_A \\ \sigma_E(Y_A) \\ 1 \end{bmatrix} = \begin{bmatrix} \tau_B \\ \sigma_E(Y_B) \\ 1 \end{bmatrix}$$

The transformation matrix $\Gamma$ is a member of a matrix Lie group (Falcone, 2017; Güngör, 2025; Stillwell, 2000).

Rosen (1995) introduced the notion of "approximate symmetry", its potential use in science, and recommended the development indices of approximate symmetry. Nugent (2024) suggested an index of approximate symmetry (IAS),

$$IAS = \sqrt{1 - \rho(Y_A)}\,,$$

where $\rho(Y_A)$ is the reliability coefficient for observed scores $Y_A$. It will now be shown that this IAS index is proportional to the standardized Euclidean distance between two distributions of observed scores $Y_A$ and $Y_B$. From above, the relationship between true scores on two measures $A$ and $B$ is given by eqn. (1),

$$\tau_B = \gamma \tau_A + \omega$$

with $\gamma \geq 1$ and $\omega$ a real number, and the relationship between the standard deviation of errors of measurement is given by eqn. (2),

$$\sigma_E(Y_B) = \gamma \cdot \sigma_E(Y_A).$$

Thus, the observed scores are:

$$Y_A = \tau_A + \epsilon_A, \text{ and}$$

$$Y_B = \gamma \tau_A + \omega + \epsilon_B,$$

where $\epsilon_A$ and $\epsilon_B$ are error scores. It follows the reliability of scores on measure A, $\rho(Y_A)$, is,

$$\rho(Y_A) = \frac{\sigma^2(\tau_A)}{\sigma^2(Y_A)}.$$

The z-scores for a person $i$ on measures $A$ and $B$ are:

$$z_{iA} = \frac{Y_{iA} - \bar{Y}_A}{\sigma(Y_A)}, \text{ and}$$

$$z_{iB} = \frac{Y_{iB} - \bar{Y}_B}{\sigma(Y_B)}.$$

The standardized Euclidean distance (SED) between the distributions of z-scores is:

$$SED = \sqrt{\frac{1}{N} \sum_{i=1}^{N} (z_{iA} - z_{iB})^2} \,.$$

Let

$$\sigma^2(Y_A) = \sigma_A^2,$$

$$\sigma^2(\tau_A) = \sigma_{\tau_A}^2, \text{ and}$$

$$\sigma^2(\epsilon_A) = \sigma_{\epsilon_A}^2.$$

Then,

$$\sigma_A^2 = \sigma_{\tau_A}^2 + \sigma_{\epsilon_A}^2, \text{ and}$$

$$\rho(Y_A) = \frac{\sigma^2(\tau_A)}{\sigma^2(Y_A)}.$$

For observed scores $Y_B$,

$$\sigma^2(\tau_B) = \gamma^2 \sigma_{\tau_A}^2, \text{ and}$$

$$\sigma^2(\epsilon_B) = \gamma^2 \sigma_{\epsilon_B}^2, \text{ and,}$$

$$\sigma^2(Y_B) = \gamma^2 \sigma_A^2.$$

Therefore, it follows that,

$$\rho(Y_B) = \rho(Y_A).$$

This is another symmetry associated with the conjunction (1) and (2).

The covariance between scores is,

$$\text{Cov}(Y_A, Y_B) = \gamma \cdot \sigma^2(\tau_A).$$

Thus, the correlation between observed scores $Y_A$ and $Y_B$ is,

$$\text{Corr}(Y_A, Y_B) = \frac{\text{Cov}(Y_A, Y_B)}{\sqrt{\sigma^2(Y_A) \cdot \sigma^2(Y_B)}} = \frac{\sigma_{\tau_A}^2}{\sigma_A^2} = \rho(Y_A) = \rho(Y_B)$$

another symmetry. Since z-scores are linear transformations of observed scores, this correlation is preserved,

$$\text{Corr}(z_A, z_B) = \rho(Y_A).$$

Now, the variance of the difference between z-scores is,

$$D^2 = \sigma^2(z_A - z_B) = \sigma^2(z_A) + \sigma^2(z_B) - 2 \cdot \text{Cov}(z_A, z_B).$$

Since both z-score distributions are standardized,

$$\sigma^2(z_A) = \sigma^2(z_B) = 1, \text{ and}$$

$$\text{Cov}(z_A, z_B) = \rho(Y_A).$$

So, it follows,

$$D^2 = 2(1 - \rho(Y_A)), \text{ so}$$

$$D = \sqrt{2(1 - \rho(Y_A))} \, .$$

Therefore, the standardized Euclidean distance between z-score distributions is proportional to Nugent's (2024) suggested IAS, with proportionality constant $\sqrt{2}$ (Lord & Novick, 1968).

## Results

### *Derivation of $\Gamma$ from Its Lie Algebra*

This section explains how the Lie group transformation $\Gamma$ arises from its Lie algebra using the exponential map, within the context of affine space. To aid understanding, a cooking metaphor is used: the ultimate "dish" to be cooked is the Lie group transformation $\Gamma$. The Lie algebra provides the directions, the exponential map gives the recipe, the associated differential equations are the specific recipe instructions, and the final solution is the finished dish, $\Gamma$.

Consider again the measurement vector (3):

$$v_A = \begin{bmatrix} \tau_A \\ \sigma_E(Y_A) \\ 1 \end{bmatrix}$$

where $\tau_A$ represents the true scores on measure $A$, and $\sigma_E(Y_A)$ represents the standard deviation of errors of measurement on measure $A$. The homogeneous coordinate - the 1 in the third row of the vector - anchors the vector in a two-dimensional affine space. A transformation group, $\Gamma$, acts on this vector to model relationships between scores on any two measures $A$ and $B$ from the k measures. From above, the transformation matrix $\Gamma$ is,

$$\Gamma = \begin{bmatrix} \gamma & 0 & \omega \\ 0 & \gamma & 0 \\ 0 & 0 & 1 \end{bmatrix}.$$

This matrix transforms the measurement vector as follows,

$$v_B = \Gamma v_A = \begin{bmatrix} \gamma & 0 & \omega \\ 0 & \gamma & 0 \\ 0 & 0 & 1 \end{bmatrix} \begin{bmatrix} \tau_A \\ \sigma_E(Y_A) \\ 1 \end{bmatrix} = \begin{bmatrix} \tau_B \\ \sigma_E(Y_B) \\ 1 \end{bmatrix}.$$

The transformation matrix $\Gamma$ is a member of a matrix Lie group (Stillwell, 2000).

*Affine Space Context*

Affine space generalizes Euclidean space by allowing transformations such as translations and scalings, but without a fixed origin. In the current context, the homogeneous coordinates enable linear algebraic operations to represent affine transformations.

*Lie Group and Lie Algebra*

Using the cooking metaphor, the final "dish," again, is the Lie group transformation acting on affine space, $\Gamma$. This represents a transformation in 2D affine space combining,

- **A uniform scaling** by $\gamma$,

- **A translation** by $\omega$, and

- **Use of a homogeneous coordinate (1)**, which enables matrix multiplication in the transformation.

The associated Lie algebra—the "recipe" for the transformation—is:

$$g = \begin{bmatrix} \ln \gamma & 0 & \frac{\omega \ln \gamma}{\gamma - 1} \\ 0 & \ln \gamma & 0 \\ 0 & 0 & 0 \end{bmatrix}.$$

This matrix encodes the infinitesimal generator of the affine transformation $\Gamma$ (Stillwell, 2000).

The Lie bracket, an important part of the Lie algebra (Stillwell, 2000), is given by the matrix equation,

$$\begin{bmatrix} \ln\gamma & 0 & 0 \\ 0 & \ln\gamma & 0 \\ 0 & 0 & 0 \end{bmatrix}\begin{bmatrix} 0 & 0 & \frac{\omega\ln\gamma}{\gamma-1} \\ 0 & 0 & 0 \\ 0 & 0 & 0 \end{bmatrix} - \begin{bmatrix} 0 & 0 & \frac{\omega\ln\gamma}{\gamma-1} \\ 0 & 0 & 0 \\ 0 & 0 & 0 \end{bmatrix}\begin{bmatrix} \ln\gamma & 0 & 0 \\ 0 & \ln\gamma & 0 \\ 0 & 0 & 0 \end{bmatrix} = \begin{bmatrix} 0 & 0 & \frac{\omega\ln\gamma}{\gamma-1} \\ 0 & 0 & 0 \\ 0 & 0 & 0 \end{bmatrix}.$$

This result shows the Lie algebra is non-commutative: the order of multiplication matters. The generators do not commute. Yet, the final term reveals a symmetry—specifically, a translation symmetry (Stillwell, 2000).

### *Step-by-Step Derivation via the Matrix Exponential*

To derive the full transformation—the "cooked dish" $\Gamma$—we compute the matrix exponential,

$$X(t) = \exp(tg) \tag{4}$$

where the parameter $t$, which ranges from 0 to 1, represents the path of the transformation. At $t = 0$, no transformation has occurred (the identity), and at $t = 1$, the transformation is complete. Given the upper triangular structure of $g$, the exponential is:

$$X(t) = \begin{bmatrix} \gamma^t & 0 & \omega \cdot \frac{\gamma^t - 1}{\gamma - 1} \\ 0 & \gamma^t & 0 \\ 0 & 0 & 1 \end{bmatrix}. \tag{5}$$

This is the one-parameter subgroup generated by $g$, describing how the affine transformation evolves from $t = 0$ to $t = 1$.

### *Differential Equations*

The evolution of $X(t)$ satisfies the matrix differential equation,

$$\frac{dX(t)}{dt} = g \cdot X(t). \tag{6}$$

From this, we extract component-wise differential equations:

- **For scaling of SD of errors of measurement**:

$$\frac{d\sigma_E(t)}{dt} = \ln \gamma \cdot \sigma_E(t), \tag{7}$$

- **For scaling and translation of true scores**,

$$\frac{d\tau(t)}{dt} = \ln \gamma \cdot \tau(t) + \frac{\omega \ln \gamma}{\gamma - 1}, \text{ and} \tag{8}$$

- **For the constant**:

$$\frac{d}{dt}(1) = 0. \tag{9}$$

These equations describe how the affine transformation evolves from infinitesimal steps—the specific instructions in the recipe for "cooking" $\Gamma$. Solving the differential equations with initial conditions $\sigma(0) = \sigma_0$ and $\tau(0) = \tau_0$, we obtain (Boyce & DiPrima, 1986),

- **For scaling of error standard deviation**:

$$\sigma_E(t) = \gamma^t \cdot \sigma_{E_0}, \tag{10}$$

- **For scaling and translation of true scores**:

$$\tau(t) = \gamma^t \cdot \tau_0 + \omega \cdot \frac{\gamma^t - 1}{\gamma - 1}, \text{ and} \tag{11}$$

- **The constant remains**,

$$1. \tag{12}$$

At $t = 1$, we obtain the original transformation group $\Gamma$,

$$\Gamma = \begin{bmatrix} \gamma & 0 & \omega \\ 0 & \gamma & 0 \\ 0 & 0 & 1 \end{bmatrix}, \tag{13}$$

(Falcone, 2017; Güngör, 2025; Olver, 1993; Stillwell, 2000).

### *Implications*

A former colleague of the author's often asked, when presented with theoretical ideas like the above, *"So what?"* He inquired about the practical significance of the theoretical notions. The above symmetries imply further symmetries. One important symmetry implied is that the population standardized mean difference (SMD), a commonly used effect size in meta-analysis (Borenstein et al., 2021; Lipsey & Wilson, 2001), will be invariant across any measures $A$ and $B$ from the $k$ measures if and only if conditions (1) and (2) above hold for the scores from measures A and B in the two populations being compared, $P_1$ and $P_2$. In other words, full measurement equivalence must hold in all subpopulations of both populations $P_1$ and $P_2$; that is, the conjunction of assumptions (1) and (2) must hold in all subpopulations of $P_1$ and $P_2$. While this follows from the symmetry principle, a proof follows.

First define:

$$\delta_A = \frac{1 - \text{rel}(Y_A)}{\text{rel}(Y_A)},$$

a term that allows direct inclusion of reliability in subsequent derivations.

Now,

$$\sigma^2(Y_A) = \sigma^2(\tau_A) + \sigma^2(E_A) = \sigma^2(\tau_A)[1 + \delta_A]$$

and

$$\sigma^2(Y_B) = \sigma^2(\tau_B)[1 + \delta_B].$$

By Theorem 2.7.2 (Lord & Novick, 1968),

$$\mu^{P_1}(\tau_A) - \mu^{P_2}(\tau_A) = \mu^{P_1}(Y_A) - \mu^{P_2}(Y_A) \text{ and}$$

$$\mu^{P_1}(\tau_B) - \mu^{P_2}(\tau_B) = \mu^{P_1}(Y_B) - \mu^{P_2}(Y_B).$$

Assumptions (1) and (2) together imply,

$$\rho(Y_A) = \rho(Y_B) \Rightarrow \delta_A = \delta_B$$

so

$$\mu^{P_1}(Y_B) - \mu^{P_2}(Y_B) = \gamma \cdot [\mu^{P_1}(Y_A) - \mu^{P_2}(Y_A)]$$

and

$$\sigma^2(\tau_B)[1 + \delta_B] = \gamma^2 \cdot \sigma^2(\tau_A)[1 + \delta_A].$$

### *Invariance of the Standardized Mean Difference*

The standardized mean difference (SMD) based on scores from measure $B$ is:

$$\text{SMD}(Y_B) = \frac{\mu^{P_1}(\tau_B) - \mu^{P_2}(\tau_B)}{\sqrt{\frac{N_{P_1}\sigma^2(\tau_B)[1+\delta_B] + N_{P_2}\sigma^2(\tau_B)[1+\delta_B]}{N_{P_1} + N_{P_2}}}}. \tag{14}$$

Substituting the transformations (1) and (2),

$$\text{SMD}(Y_B) = \frac{\gamma \cdot [\mu^{P_1}(\tau_A) - \mu^{P_2}(\tau_A)]}{\gamma \cdot \sqrt{\frac{N_{P_1}\sigma^2(\tau_A)[1+\delta_A] + N_{P_2}\sigma^2(\tau_A)[1+\delta_A]}{N_{P_1} + N_{P_2}}}} = \text{SMD}(Y_A) \tag{15}$$

Thus, the population SMD remains invariant across measures $A$ and $B$ from the k measures under the specified measurement equivalence conditions, i.e., the conjunction of assumptions (1) and (2) in both $P_1$ and $P_2$.

### *Invariance of Population SMD Implies Assumptions (1) and (2)*

Now assume that the population SMD is invariant across scores from measures A and B,

$$\text{SMD}(Y_A) = \text{SMD}(Y_B). \tag{16}$$

For this equality to hold, both the numerator and denominator of the SMD must be scaled by the same factor $\gamma \geq 0$. That is:

$$\text{SMD}(Y_B) = \frac{\gamma \cdot [\mu^{P_1}(\tau_A) - \mu^{P_2}(\tau_A)]}{\gamma \cdot \sqrt{\frac{N_{P_1}[\sigma^2(\tau_A(t))[1+\delta_A]]^{P_1} + N_{P_2}[\sigma^2(\tau_A(t))[1+\delta_A]]^{P_2}}{N_{P_1} + N_{P_2}}}} = \text{SMD}(Y_A) \tag{17}$$

which represents a uniform scaling of both numerator and denominator by $\gamma$. This is a symmetry transformation (Rosen, 1995).

Additionally, a constant translation $\omega$ may be added to both group means in the numerator,

$$[\gamma \cdot \mu^{P_1}(\tau_A) + \omega] - [\gamma \cdot \mu^{P_2}(\tau_A) + \omega] = \mu^{P_1}(\tau_B) - \mu^{P_2}(\tau_B). \tag{18}$$

Since $\omega$ cancels out, the difference remains unchanged. Importantly, adding a constant does not affect the denominator, because the variance of a variable plus a constant equals the variance of the original variable (Lord & Novick, 1968).

Therefore, for the SMD to remain invariant across transformed measures, the following conditions must hold:

- **Uniform scaling of true scores by γ and possible translation by ω**:

$$\tau_B = \gamma \cdot \tau_A + \omega \tag{19}$$

- **And uniform scaling of SD of measurement error by γ**:

$$\sigma_E(Y_B) = \gamma \cdot \sigma_E(Y_A). \tag{20}$$

Thus, the population SMD will be invariant across the scores from the $k$ measures if and only if the conjunction of these transformation conditions is satisfied.

### *Dynamic Transformation via Differential Equations*

The invariance of the standardized mean difference (SMD) can also be demonstrated from a Lie algebra perspective. Assume scores for both populations evolve under the same transformation flow from measure $A$ to measure $B$ governed by the Lie algebra of the transformation matrix $\Gamma$. The differential equations derived from the Lie algebra are,

$$\frac{d\tau_A(t)}{dt} = \ln(\gamma) \cdot \tau_A(t) + \frac{\omega \ln(\gamma)}{\gamma - 1} \text{ , and} \tag{21}$$

$$\frac{d\sigma_E(t)}{dt} = \ln(\gamma) \cdot \sigma_E(t). \tag{22}$$

Assuming the population means $\mu^{P_1}(t)$ and $\mu^{P_2}(t)$ follow the same form as $\tau_A(t)$, we have,

$$\frac{d\mu^{P_1}(t)}{dt} = \ln(\gamma) \cdot \mu^{P_1}(t) + \frac{\omega \ln(\gamma)}{\gamma - 1} \text{, and} \tag{23}$$

$$\frac{d\mu^{P_2}(t)}{dt} = \ln(\gamma) \cdot \mu^{P_2}(t) + \frac{\omega \ln(\gamma)}{\gamma - 1}. \tag{24}$$

Define the standardized mean difference as a function of the transformation parameter $t$ as,

$$\text{SMD}(t) = \frac{\mu^{P_1}(t) - \mu^{P_2}(t)}{\sqrt{\sigma^2_{\text{pooled}}(t)}}, \tag{25}$$

where the pooled variance is,

$$\sigma^2_{\text{pooled}}(t) = \frac{N_{P_1}[\sigma^2(\tau_A(t))[1+\delta_A]]^{P_1} + N_{P_2}[\sigma^2(\tau_A(t))[1+\delta_A]]^{P_2}}{N_{P_1} + N_{P_2}}. \tag{26}$$

Differentiating $\text{SMD}(t)$ using the quotient rule,

$$\frac{d}{dt}\text{SMD}(t) = \frac{\sqrt{\sigma^2_{\text{pooled}}(t)} \cdot (\frac{d\mu^{P_1}}{dt} - \frac{d\mu^{P_2}}{dt}) - (\mu^{P_1}(t) - \mu^{P_2}(t)) \cdot \frac{d}{dt}\sqrt{\sigma^2_{\text{pooled}}(t)}}{\sigma^2_{\text{pooled}}(t)}. \tag{27}$$

Substituting the differential equations:

$$\frac{d\mu^{P_1}}{dt} - \frac{d\mu^{P_2}}{dt} = \ln(\gamma) \cdot (\mu^{P_1}(t) - \mu^{P_2}(t)), \text{ and} \tag{28}$$

$$\frac{d}{dt}\sqrt{\sigma^2_{\text{pooled}}(t)} = \ln(\gamma) \cdot \sqrt{\sigma^2_{\text{pooled}}(t)}, \tag{29}$$

we obtain,

$$\frac{d}{dt}\text{SMD}(t) = 0. \tag{30}$$

Thus, the standardized mean difference remains invariant under the transformation flow generated by the Lie algebra—provided the scores are transformed identically in both populations under the same $\gamma > 0$ and $\omega$.

### *Reverse Implication*

Assume the population standardized mean difference *SMD*(t) is invariant under a transformation flow whenever the numerator difference and the pooled standard deviation evolve with the same instantaneous multiplicative rate. The Lie algebra of the transformation matrix $\Gamma$ corresponds to the special case where this rate is constant, $\alpha(t) = \ln(\gamma)$, specifically:

$$\frac{d\tau_A(t)}{dt} = \ln(\gamma)\,\tau_A(t) + \frac{\omega\ln(\gamma)}{\gamma - 1}. \qquad (31)$$

If the transformation flow deviates from this condition—so that the numerator and denominator no longer share the same rate of change—the SMD will not remain invariant across the transformation.

The standardized mean difference at transformation parameter $t$ is,

$$\text{SMD}(t) = \frac{\mu^{P_1}(t) - \mu^{P_2}(t)}{\sqrt{\sigma^2_{\text{pooled}}(t)}} \qquad (32)$$

For simplicity, assume equal population sizes, so,

$$\sigma^2_{\text{pooled}}(t) = \frac{\left[\sigma^2(\tau_A(t))[1+\delta_A]\right]^{P_1} + \left[\sigma^2(\tau_A(t))[1+\delta_A]\right]^{P_2}}{2}, \qquad (33)$$

Suppose the transformation flow is governed by differential equations that differ from the Lie algebra–generated form. That is, assume:

$$\frac{d\mu^{P_1}}{dt} = f_1(\mu^{P_1}(t), t), \qquad (34)$$

$$\frac{d\mu^{P_2}}{dt} = f_2(\mu^{P_2}(t), t), \qquad (35)$$

$$\frac{d\sigma^2(E_A^{P_1}(t))}{dt} = g_1(\sigma^2(E_A^{P_1}(t)), t), \text{ and} \qquad (36)$$

$$\frac{d\sigma^2(E_A^{P_2}(t))}{dt} = g_2(\sigma^2(E_A^{P_2}(t)), t). \qquad (37)$$

Then the derivative of SMD$(t)$ is:

$$\frac{d}{dt}\text{SMD}(t) = \frac{\sqrt{\sigma^2_{\text{pooled}}(t)} \cdot [f_1(\mu^{P_1}(t),t) - f_2(\mu^{P_2}(t),t)] - (\mu^{P_1}(t) - \mu^{P_2}(t)) \cdot \frac{d}{dt}\sqrt{\sigma^2_{\text{pooled}}(t)}}{\sigma^2_{\text{pooled}}(t)} \qquad (38)$$

Unless both of the following conditions are satisfied,

$$f_1(\mu^{P_1}(t),t) - f_2(\mu^{P_2}(t),t) = \ln(\gamma) \cdot (\mu^{P_1}(t) - \mu^{P_2}(t)), \text{ and} \qquad (39)$$

$$\frac{d}{dt}\sqrt{\sigma^2_{\text{pooled}}(t)} = \ln(\gamma) \cdot \sqrt{\sigma^2_{\text{pooled}}(t)}, \qquad (40)$$

the numerator will not cancel to zero, and therefore the population SMD will not be invariant across the measures $A$ and $B$.

### Effects of Breaking the Symmetry Expressed by (1) and (2)

The conjunction of conditions (1) and (2) represents a symmetry of uniform scaling of true scores and error standard deviations by γ, along with the translation symmetry of true scores by $\omega$. We now consider an important implication of the above explicated measurement model: the consequences of breaking this symmetry (1) and (2). First note that assumption (1) can be expressed as:

$$\tau_B = \gamma\tau_A^1 + \omega \qquad (1A)$$

a special case of the more general expression:

$$\tau_B = \gamma\tau_A^k + \omega, \qquad (2A)$$

with k > 0.

This general expression introduces nonlinearity into the relationship between true scores. It is used in the simulation study below to examine the effects of small deviations from the symmetry, expressed by (1), on the population true-score-level SMD effect size.

For simplicity, throughout the simulation we assume $\gamma = 1$, $\omega = 0$, and no measurement error. This allows us to focus on the true score SMD, representing the true score effect size. We begin with the assumption that the relationship between the true scores on measures $A$ and $B$ is:

$$\tau_B = 1 \cdot \tau_A + 0 = \tau_A$$

in both populations $P_1$ and $P_2$. In this case, as shown above, the population true score SMD will be the same whether based on scores from measure $A$ or measure $B$ (Falcone, 2017; Güngör, 2025; Olver, 1993; Stillwell, 2000).

### Simulation Study

Populations of true scores were created for populations $P_1$ and $P_2$, each with 1,000,000 scores. The mean of $P_1$ was 63.05 ($SD = 13.08$), and the mean of $P_2$ was 63.04 ($SD = 13.06$). The relationship between these true scores was:

$$\tau_B = 1.125\tau_A$$

in both $P_1$ and $P_2$. Next, varying degrees of lack of measurement equivalence were introduced in scores for population $P_2$ by varying $k$ in (1b) from $k = 1.000$ to $k = 1.020$ in steps of $\Delta k = 0.001$. For example, in one simulation step the relationship between true scores in $P_1$ was:

$$\tau_B = 1 \cdot \tau_A = \tau_A$$

while in $P_2$ it was:

$$\tau_B = \tau_A^{1.001}$$

The $\omega$ term was not included in the simulation since it cancels out in the numerator of the SMD computation.

In each scenario across the simulations, $k = 1.000$ in $P_1$, while $k$ varied slightly above 1.000 in $P_2$, representing a small lack of measurement equivalence. At each step, the population true score SMD comparing $P_1$ and $P_2$ was computed. The results are plotted in Figure 1.

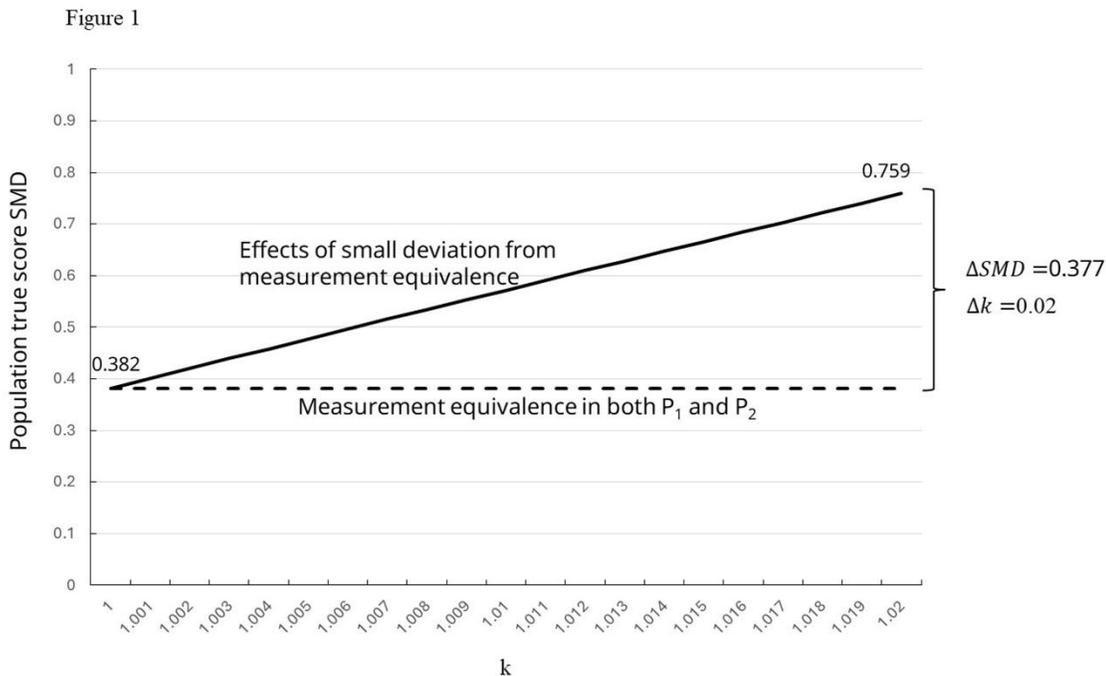

Figure 1: Results of simulation. The dashed line shows the population true score SMD when measurement equivalence holds in both populations. The solid line shows how the population true score SMD deviates from invariance as symmetry condition (1) is increasingly broken.

In the Figure, the dashed line represents the case in which full symmetry holds, while the solid line shows how the population true score SMD changes as the deviation from symmetry increases. As shown in the figure, as the symmetry condition $k = 1.000$ deviates further from perfect symmetry, the population SMD deviates further from invariance. As $k$ increases above 1.000, the population SMD deviates further from invariance.

## Discussion

This paper has extended Nugent's (2024) matrix Lie group measurement model by identifying additional symmetries and demonstrating their implications for effect size invariance in meta-analysis. Specifically, it was shown that the population true score standardized mean difference (SMD) remains invariant across different measures if and only if the transformations between those measures satisfy two key conditions:

- A uniform linear scaling of true scores involving both a uniform scaling by γ and a translation, and

- A uniform scaling of measurement error by γ.

These conditions define a symmetry group whose structure is captured by a Lie algebra, and whose global behavior is governed by the exponential map in affine space.

By deriving the transformation flow from the Lie algebra and solving the associated differential equations, it was shown that the SMD remains constant along this path—confirming that the invariance of the SMD effect size is not merely a statistical artifact but a geometric consequence of the measurement symmetry. Conversely, any deviation from this transformation flow results in a non-zero derivative of the SMD indicating a breakdown of invariance.

One of the symmetries identified by Nugent (2024) is the invariance of the relative ordering of scores in the true scores from measures A and B if the conjunction (1) and (2) holds for the scores on the k measures. This invariance is manifested by the true z-scores for true scores on measures A and B. It is this relative ordering that conveys information about the concept measured into data analyses. Thus, if the conjunction of (1) and (2) holds, in a sense the information contained in the z-score distribution is conserved across the k measures. This parallels Noether's theorem which states, in general, that symmetries are associated with conserved quantities. Thus, analogously, the symmetries associated with the combination of transformations (1) and (2) are associated with conservation of information conveyed by true z-scores from the k measures. This conservation has important implications for validity.

An implication of the results concerning the invariance of the population true score SMD is that meta-analysts should consider evidence of measurement equivalence prior to assuming scores based on different measures are comparable in a meta-analysis. Lack of measurement equivalence would have implications not only in terms of direct comparability of score meaning, but also what differences in scores indicate, lack of measurement equivalence of differences in the construct inferred from the scores.

The simulation study illustrated how even infinitesimal violations of the linear symmetry condition—modeled by a nonlinear exponent k slightly deviating from 1—can lead to distortions in the population true score SMD. This finding underscores the sensitivity of effect size comparability to the underlying measurement structure and highlights the importance of verifying measurement equivalence when synthesizing results across studies.

It is also important to note that while uniform scaling of measurement error is a symmetry condition that guarantees invariance at the true score level, at the observed score level it attenuates effect sizes in a predictable manner. Specifically, observed SMDs are reduced relative to true SMDs in proportion to the reliability of the measures,

$$SMD_{observed} = \sqrt{\rho} \cdot SMD_{true}. \qquad (41)$$

This predictability is a consequence of symmetry. Thus, invariance at the latent construct level coexists with systematic attenuation at the observed level, underscoring the importance of considering measurement error when interpreting effect sizes.

These findings also have important implications for longitudinal research. Invariance under transformation flows implies that effect sizes can remain stable over time only if measurement conditions are symmetric across repeated waves. Even small deviations from symmetry can mimic or mask genuine developmental change, creating the appearance of growth or decline where none exists. This underscores the need for formal measurement equivalence testing across time points to ensure that observed changes in standardized mean differences reflect true developmental or treatment effects rather than artifacts of measurement structure.

Together, these results provide a mathematical foundation for understanding when and why effect sizes such as the SMD can be meaningfully compared across instruments and across time. They also offer a novel application of Lie group theory to psychometrics, opening new avenues for exploring symmetry, invariance, and transformation in measurement models. Similarly, in cross-cultural or cross-language research, ensuring that measurement transformations preserve symmetry is essential for meaningful comparison of effect sizes across populations.

At its core, this analysis is an application of the symmetry principle: the idea that symmetric structures yield symmetric outcomes. When measurement transformations, population sizes, and error variances are symmetrically aligned across groups, the population standardized mean difference remains invariant. This invariance is not coincidental—it emerges naturally from the underlying symmetries encoded in the Lie group transformation. Conversely, when these symmetries are broken, even infinitesimally, the SMD begins to deviate, revealing the sensitivity of statistical outcomes to structural asymmetries. Thus, the symmetry principle provides a unifying lens through which measurement equivalence, effect size comparability, and transformation invariance can be understood and explored. The symmetry principle also extends to measurement error: while uniform scaling preserves invariance at the true score level, it produces predictable attenuation at the observed score level, reinforcing that both invariance and attenuation are structured consequences of the same underlying symmetries.